\documentclass[aps,pra,showpacs,twoside,10pt,floatfix,nofootinbib,longbibliography,twocolumn]{revtex4-2}
\usepackage[colorlinks=true, citecolor=red, urlcolor=blue ]{hyperref}
\usepackage{times,epsfig,amssymb,amsfonts,amsmath,bm,subfigure,mathtools,amsthm,braket,soul,enumitem,xcolor,physics,graphics,graphicx}

\usepackage[normalem]{ulem}
\usepackage{comment, appendix}
\usepackage{epstopdf}

\usepackage{comment}

\begin{document}

\title{
Harnessing non-Hermiticity for efficient quantum state transfer
}

\author{Sejal Ahuja, Keshav Das Agarwal, Aditi Sen (De)}
\affiliation{Harish-Chandra Research Institute, A CI of Homi Bhabha National Institute,  Chhatnag Road, Jhunsi, Allahabad - 211019, India}

\begin{abstract}
The non-Hermitian Hamiltonian describes the effective dynamics of a system coupled to a continuously measured bath, and  
can exhibit anti-unitary symmetries that give rise to exceptional points and broken phases with complex eigenvalues, features unique to non-Hermitian systems. Going beyond conventional Hermitian physics, we analyze the impact of  non-Hermiticity in the quantum state transmission by employing a non-Hermitian spin chain that functions as a quantum data bus. By deriving a general expression for the fidelity of quantum state transfer for a \(U(1)\)-symmetric non-Hermitian Hamiltonian, we analyze 
$\hat{\mathcal{P}}\hat{\mathcal{T}}$-symmetric \(XX\) and SSH models, complemented by a numerical study of the 
$\hat{\mathcal{R}}\hat{\mathcal{T}}$-symmetric \(iXY\) model. We demonstrate that, in several parameter regimes, the transfer fidelity in the non-Hermitian setting exceeds the classical threshold and can even exceed the performance of the corresponding Hermitian models. In particular, for the SSH model with dominant inter-cell coupling, the broken phase supports near-unit-fidelity quantum state transfer, a level of performance that the corresponding Hermitian model fails to attain. Moreover, we establish a correspondence between the non-Hermitian and Hermitian descriptions by identifying related parameter regions in which the fidelity fails to surpass the classical bound.

\end{abstract}

\maketitle

\section{Introduction}

Protecting quantum technologies against decoherence is a key challenge in demonstrating their benefit over existing classical devices. This stimulates the development of quantum machines whose performance remains stable despite inevitable interactions with the environment. In this context, non-Hermitian models play a crucial role, as they naturally arise in effective descriptions of open quantum systems. Here, system–environment interactions are captured by Lindblad operators representing quantum jumps, and the dynamics is governed by the Gorini–Kossakowski–Lindblad–Sudarshan (GKLS) master equation~\cite{open_quan_book, lidar_2020_lecture}. 
The non-Hermitian Hamiltonians describe post-selected dynamics, corresponding to trajectories without quantum jumps during system–bath interactions~\cite{open_quan_book, lidar_2020_lecture, Minganti2020}. Such regimes naturally arise in gain–loss systems~\cite{Eleuch2015}  and in continuously monitored environments, where weak measurements effectively condition the system’s evolution~\cite{turkeshi_prb_2021, fleckenstein_prr_2022, turkeshi_prb_2023b}. 
Beyond standard Hermitian quantum mechanics, non-Hermitian models incorporate anti-unitary symmetries, giving rise to unbroken (with symmetric eigenstates) and broken phases (with eigenstates breaking the symmetry) separated by exceptional points~\cite{bender_spectra, mosta_ali_2002, Mostafazadeh_2004, Song_RT_symm, Jin2009, Joglekar2010, kda2024}. 
Consequently, they exhibit non-trivial phases~\cite{Gong2018, Kawabata2019, Bergholtz2021}, rich entanglement properties~\cite{turkeshi_prb_2023b, Kawabata2023, Agarwal2023May} and non-Hermitian skin effect~\cite{Li2020Oct, xiao_prb_2022} with no Hermitian counterpart. Remarkably, non-Hermitian models have been shown to improve the performance of various quantum technologies, including quantum thermal machines~\cite{khandelwal_prx_2021, konar2022quantum, Santos2023Dec} and quantum sensing~\cite{ep_sensing_1, ep_sensing_2, ep_sensing_3, agarwal2025} and have recently been  realized experimentally  under suitable conditions~\cite{guo_pt_exp, chitsazi_prl_2017, naghiloo_nature_ep_sensing_2019, ep_sensing_exp_2}.

\begin{figure}
    \centering
\includegraphics[width=\linewidth]{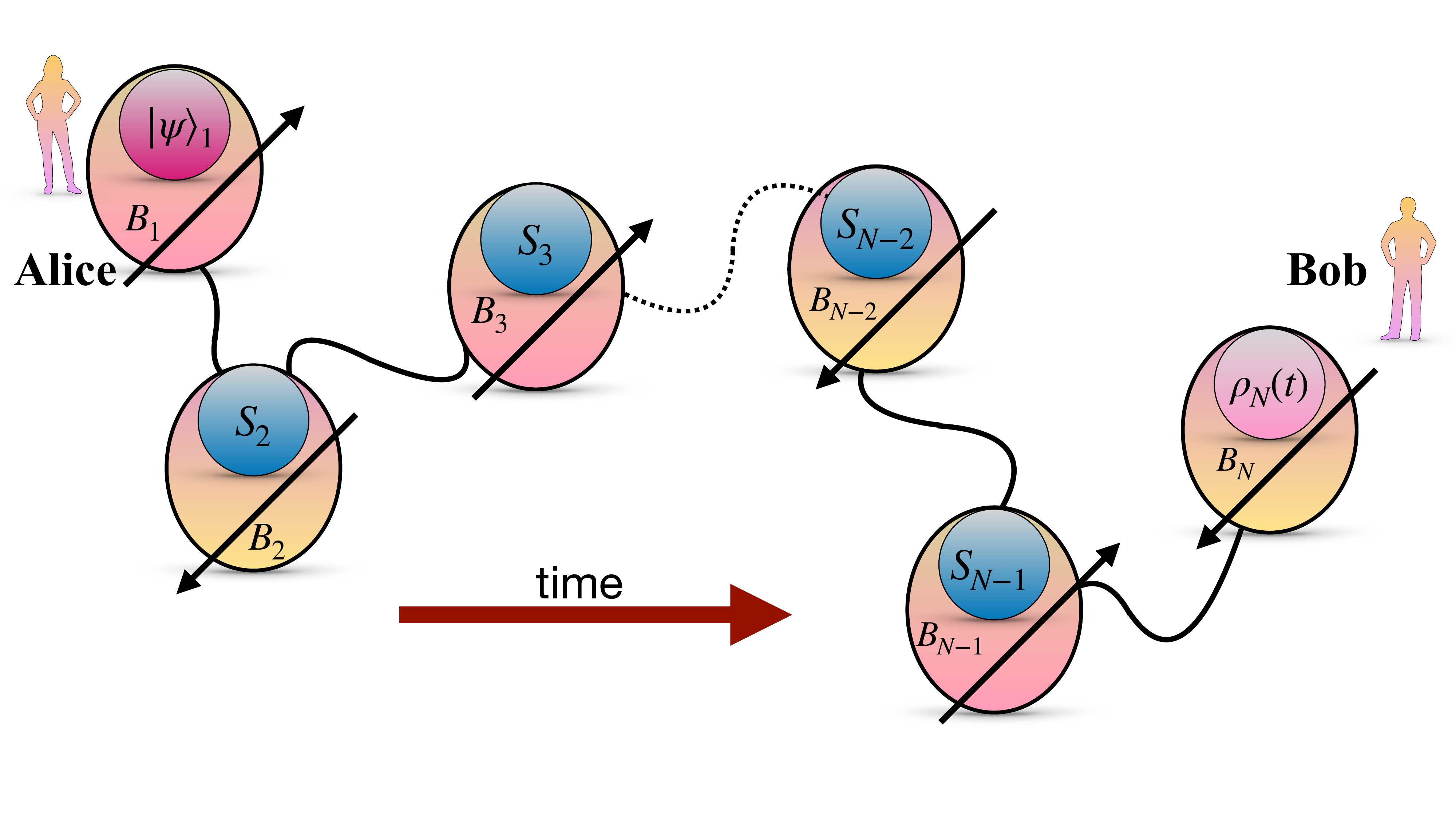}
    \caption{
    {\bf Schematic of the QST protocol through a spin chain.} Each spin is coupled to a local bath \(B_k\) (\(k=1, 2, \ldots N\)). Owing to the presence of these baths and the suitable continuous measurements followed by post-selection, the state-transfer dynamics is governed by an effective $\hat{\mathcal{P}}\hat{\mathcal{T}}$-symmetric non-Hermitian Hamiltonian. In the protocol, \(|\psi\rangle_1\) denotes the arbitrary quantum state to be transferred, (\(S_2,S_3,...,S_{N-1}\)) represent the intermediate spins of the chain and  \(\rho_N(t)\) is the final state received at Bob’s site after time evolution. By appropriately selecting the initial state, the governing Hamiltonian, and its parameters, our goal is to maximize the fidelity, which serves as a measure of the protocol’s success. }
    \label{fig:sch}
\end{figure}

A crucial requirement for many quantum devices including building a scalable quantum computer is the establishment of a robust quantum link that enables the transfer of quantum states with the highest possible fidelity. The seminal work by Bennett et al.~\cite{BBCJPWtele} demonstrated that unit-fidelity transfer can be achieved using a shared maximally entangled state between the sender (Alice) and the receiver (Bob), followed by a measurement on Alice’s side. Alternatively, quantum state transfer (QST) can also be realized using initially separable states that evolve under an interacting Hamiltonian~\cite{Bose2003, Bose2007}. During this evolution, entanglement is dynamically generated between Alice and Bob, allowing high-fidelity state transfer that is unattainable with unentangled states~\cite{zukoski_pra_2005, massar_95}. In uniform Heisenberg spin chains, the transfer fidelity typically decreases with increasing system size; however, perfect QST  can be achieved by suitably engineering site-dependent interaction strengths~\cite{Christandl2004, Osborne2004, Christandl2005, Burgarth2005, kay2010}. QST protocols have been extensively investigated in a wide range of many-body systems~\cite{Li2005, Giovannetti2006, Chen2007}, including spin-1 chains~\cite{Romero2007}, multiple channels~\cite{Yousefjani2020}, continuous variable systems~\cite{Xu2023, Anuradha2025}, and long-range interacting models~\cite{Gualdi2008, Gualdi2009, Eldredge2017, Hermes2020, Hong2021, Tran2021, Nikolaos2023, qst_lr}.

In realistic implementations, the system inevitably interacts with its environment, which motivates the study of QST within the framework of non-Hermitian models, an aspect that remains relatively underexplored (see recent studies on non-unitary quantum walks~\cite{Salimi2010, Badhani2024, Acuaviva2025} and perfect QST with engineered \(\hat{\mathcal{P}}\hat{\mathcal{T}}\)-symmetric systems~\cite{Zhang2012}). 
In this work, we eliminate this gap by investigating how non-Hermiticity arising from system–environment interactions influences quantum state transfer protocols (see Fig. \ref{fig:sch} for schematics). Starting from an initial product state, we derive an analytical expression of the transfer fidelity in a generic \(U(1)\)-symmetric non-Hermitian Hamiltonian. We then demonstrate QST under dynamics governed by $\hat{\mathcal{P}}\hat{\mathcal{T}}$-symmetric Hamiltonian, namely the \(XX\)~\cite{barouch_pra_1970, barouch_pra_1971} and Su–Schrieffer–Heeger (SSH) models \cite{ssh_og, Asboth2016, Lieu2018, Halder_2023}, in the presence of a uniform magnetic field together with an imaginary alternate magnetic field, which effectively arises from coupling to a continuously monitored environment. Note that unlike Hermitian Hamiltonians, where QST proceeds through completely positive and trace-preserving (CPTP) channels, non-Hermitian dynamics give rise to non-CPTP channels.

We identify parameter regimes in which the non-Hermitian models outperform their Hermitian counterparts in terms of  fidelity. To systematically characterize these regimes, we focus on a key figure of merit -- the first maximum of the fidelity that surpasses the classical threshold 
\(\mathcal{F} > \mathcal{F}_c=2/3\) \cite{massar_95, tele_fid99}, thereby ensuring quantum benefit. Our analysis reveals that the quantum advantage provided by non-Hermiticity is strongly tied to the symmetry of the eigenspectrum, namely, whether the system lies in the unbroken or broken  regime, as well as to the structure of the underlying Hermitian component of the Hamiltonian.
In particular, for the \(XX\) model, enhanced fidelity is achieved in the unbroken regime, which is associated with substantial entanglement generation during the evolution. We further uncover a {\it correspondence} between parameter sets in the Hermitian model and those in the unbroken regime of the non-Hermitian model where the fidelity fails to exceed the classical bound. This correspondence also appears in the SSH model when the intra-cell coupling dominates. Conversely, when the inter-cell coupling is stronger, the broken regime yields high-fidelity QST that improves with increasing inter-cell strength, while the corresponding Hermitian model exhibits a shrinking parameter region surpassing the classical limit. Moreover, by extending our study to the \(\mathcal{RT}\)-symmetric
\(iXY\) model~\cite{Song_RT_symm, gan_adi_factor, Agarwal2023May, kda2024}, we further demonstrate that non-Hermiticity can enhance QST performance, underscoring its constructive role in quantum information transfer.

The paper is organized as follows. Sec. \ref{sec:framework} introduces the QST protocol, its figure of merit, the non-Hermitian model considered in this paper and  the derivation of fidelity for a generic class of $U(1)$-symmetric non-Hermitian evolution Hamiltonians. We  establish the benefit of non-Hermiticity in QST for the $XX$ model in Sec. \ref{sec:xx_model}, for SSH model in Sec. \ref{sec:ssh_model} and \(\hat{\mathcal{R}}\hat{\mathcal{T}}\)-symmetric $iXY$ model in Sec. \ref{sec:ixy_model}. The results are summarized in Sec. \ref{sec:conclu}.


\begin{figure*}
    \centering
\includegraphics[width=\linewidth]{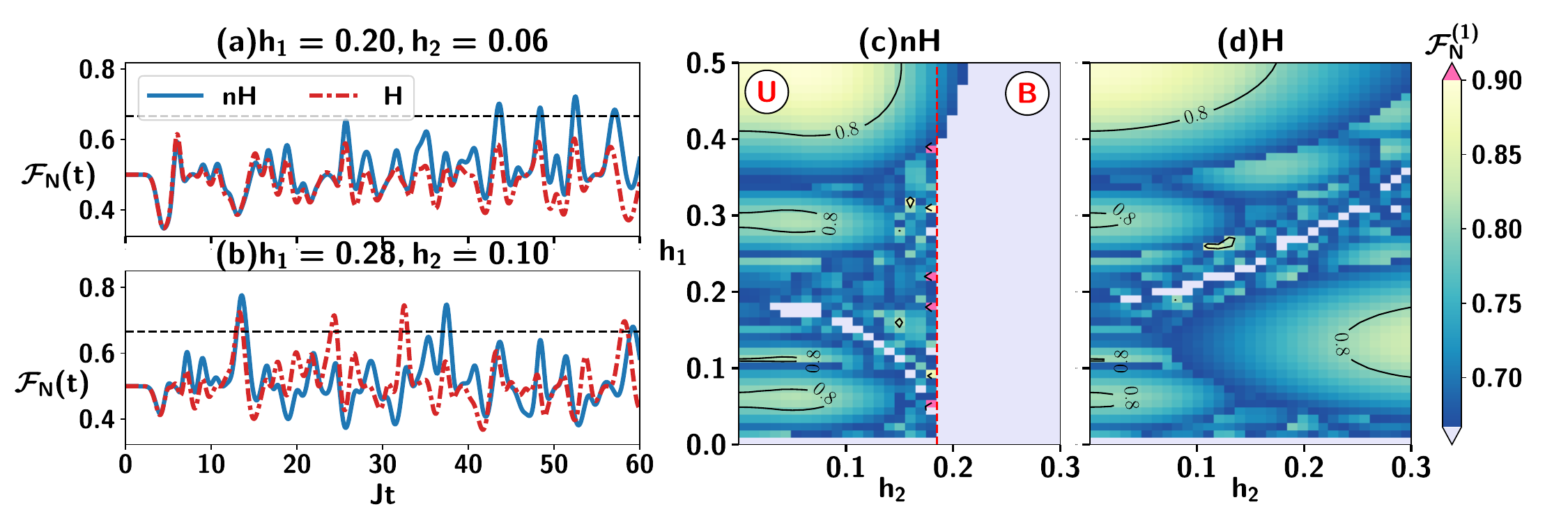}
    \caption{{\bf Comparison between fidelities \(\mathcal{F}_N(t)\) of QST for the non-Hermitian and Hermitian $XX$ models.} (a) and (b) \(\mathcal{F}_N(t)\) (ordinate) against \(t\) (abscissa) for specific choices of \((h_1,h_2)\), with advantage in the nH case, i.e., \(\mathcal{F}_{N, nH}^{(1)}>\mathcal{F}_{N, H}^{(1)}\). Here, solid (blue) lines indicate non-Hermitian case while dashed-dot (red) lines show the corresponding Hermitian behavior, with dashed (black) line denoting the classical limit $\mathcal{F}_c$. The fidelity corresponding to non-Hermitian evolution beats the classical limit of \(2/3\) earlier, and the first maxima rises above the corresponding Hermitian evolution, where \(h_1\) and \(h_2\) belong to the unbroken regime of the non-Hermitian Hamiltonian in both cases. The map plot of first maximum fidelity $\mathcal{F}_N^{(1)}$ in the \((h_2,h_1)\)-plane for (c) non-Hermitian, and (d) Hermitian \(XX\) models. The points where fidelity rises above \(0.9\) are shown in pink. The red dotted line in (c) distinguishes the broken (B) region (\(h_2 \gtrsim 0.18\)) from the unbroken (U) one. The parameters in the unbroken region of (c) and (d) represent an ellipse and a hyperbola, respectively, of the points where the fidelity could not reach the classical threshold. The corresponding parameters of these curves in Eq. (\ref{eq:ellipse}) and (\ref{eq:hyperbola}) are (\(a=0.188, b=0.186\)). Here \(N=16\) and all the axes are dimensionless.}
    \label{fig:plt_xx_ub}
\end{figure*}

\section{Framework for non-hermitian Quantum state transfer }
\label{sec:framework}

{\it Generic set-up for state transfer.}
Suppose a sender, say, Alice, situated at site \(1\), aims to transfer an arbitrary state, \(|\psi\rangle = \cos\frac{\theta}{2}|0\rangle + e^{i\phi}\sin\frac{\theta}{2} |1\rangle\) (\(0 \leq \theta \leq \pi\) and \(0 \leq \phi \leq 2\pi\)), to a receiver, say, Bob, located at site \(N\). Alice and Bob are connected through a quantum spin chain consisting of (\(N-2\)) spin-\(1/2\) particles, which act as a quantum channel for the state transfer (see Fig. \ref{fig:sch} for schematics).

Initially, all these (\(N-1\)) spins (including Bob) are prepared as  \(\bigotimes_{j=2}^{N}|0\rangle_{j}\). Hence, the initial state of the system of \(N\) qubits reads as  \vspace{-2mm}
\begin{equation}
    \ket{\Psi(0)} \!=\! \ket{\psi}_1 \!\otimes\! \bigotimes\limits_{j=2}^{N}\!|0\rangle_{j} \!=\! \cos\frac{\theta}{2}\left(\bigotimes\limits_{j=1}^{N}\!\ket{0}_{j}\right) \!+\! e^{i\phi}\sin\frac{\theta}{2} \ket{\mathbf{1}},\nonumber \vspace{-3mm}
\end{equation}
where $\ket{\mathbf{k}} \!=\! \ket{1}_k\!\!\bigotimes\limits_{j=1,j\neq k}^{N}\!\ket{0}_{j}$ are $N$ orthogonal states with $k\!=\!1$ $,...,$ $N$.


In order to transmit the state from Alice at  site \(1\) to Bob at site \(N\), the system evolves under an interacting Hamiltonian $\hat{\mathcal{H}}$, with $\ket{\Psi(t)}\!=\!e^{-i\hat{\mathcal{H}}t}\ket{\Psi(0)}$ as the time evolved state. Note that the evolved state has to be normalized at each time in the case of the non-Hermitian evolution. The state at Bob's site is $\rho_N(t)\!=\!\text{Tr}_{\bar{N}}(\ket{\Psi(t)}\!\bra{\Psi(t)})$, where $\text{Tr}_{\bar{k}}(\cdot)$ is the partial trace over all sites except $k$. Therefore, the fidelity $\mathcal{F}_{\ket{\psi},N}(t) \!=\! \bra{\psi}\rho_{N}(t)\ket{\psi} \!\in\![0,1]$ gives a benchmark of the quantum state transfer protocol with $\mathcal{F}_{\ket{\psi},N}(t)\!\sim\!1$ indicating perfect QST of $\ket{\psi}$. To make the indicator state-independent, we define the average fidelity,  
\begin{equation}
    \mathcal{F}_N(t) \!=\! \int_{\text{Haar}}\!\!\!\!\! d\ket{\psi}\mathcal{F}_{\ket{\psi},N}(t) \!=\! \int_{\text{Haar}}\!\!\!\! \!d\ket{\psi}\bra{\psi}\rho_{N}(t)\ket{\psi},
\end{equation}
where $N$ and $t$ are the parameters of the protocol, and ``Haar" written in the integration indicates that the integration has to be performed over the entire state space, i.e., over the Haar uniformly generated states \cite{zukoski_pra_2005,Mele2024}. We call the average fidelity (in short, fidelity now onward) capable of providing quantum gain, when $\mathcal{F}_N(t)\!>\!\mathcal{F}_c \!=\! 2/3$, where $\mathcal{F}_c$ is known as the classical fidelity~\cite{massar_95, tele_fid99}.
In this context, we compute the first maximum value of fidelity, \(\mathcal{F}_N^{(1)}\) which is above \(\mathcal{F}_c\), occurred at time \(t\)  to assess the beneficial role of the evolving Hamiltonian  for quantum state transmission. Note that it is plausible to assume that  the fidelity \(\mathcal{F}_N^{(1)}\) goes beyond \(\mathcal{F}_c\) occurred within finite time \(t^*\) which is fixed to \(t^*=200\) in our study.

{\it Non-Hermitian and Hermitian models.} After the preparation of the initial state, the dynamics, in our case, is governed by the $\hat{\mathcal{P}}\hat{\mathcal{T}}$- or $\hat{\mathcal{R}}\hat{\mathcal{T}}$-symmetric non-Hermitian (nH) Hamiltonian (Appendix \ref{sec:appen_def}). Our goal is to identify the situation where non-Hermitian system provides benefit over the Hermitian (H) counterpart. Specifically, we consider two scenarios where $\hat{\mathcal{H}}\!=\hat{\mathcal{H}}_{H},\hat{\mathcal{H}}_{nH}$, with
\begin{eqnarray}
\hat{\mathcal{H}}_H = \hat{\mathcal{H}}_1 + \hat{\mathcal{H}}_2, \quad \hat{\mathcal{H}}_{nH} = \hat{\mathcal{H}}_1 + i\hat{\mathcal{H}}_2,
\label{eq:nhH_def}
\end{eqnarray}
where \(\hat{\mathcal{H}}_1\) and \(\hat{\mathcal{H}}_2\) are both individually Hermitian. When the nH Hamiltonian $\hat{\mathcal{H}}_{nH}$ possesses a complete real eigenspectrum for certain parameters, it is known as pseudo-Hermitian regime as both $\hat{\mathcal{H}}_{nH}$ and $\hat{\mathcal{H}}_{H}$ are connected by a similarity transformation~\cite{mosta_ali_2002, Mostafazadeh_2004}. Typically by tuning system parameters, the system undergoes a transition from a complete real eigenspectrum (unbroken) to a complex one (broken) which occurs at the point, known as exceptional point \cite{Ashida2020, Bergholtz2021}.

For the $\hat{\mathcal{P}}\hat{\mathcal{T}}$-symmetric nH Hamiltonian, we consider \(\hat{\mathcal{H}}_1\) and \(\hat{\mathcal{H}}_2\) with $U(1)$-symmetry, where \(\hat{\mathcal{H}}_2\) represents the local alternating transverse magnetic field. In particular, 
\begin{align}
 \nonumber \hat{\mathcal{H}}_1 &\!=\! \!\sum_{k=1}^{N-1} J_k (\hat{\sigma}^x_k\hat{\sigma}^x_{k+1} \!+\! \hat{\sigma}^y_k\hat{\sigma}^y_{k+1}) + h_1^\prime\sum_{k=1}^N \hat{\sigma}_k^z, \\ 
\hat{\mathcal{H}}_2 &\!=\! h_2^\prime\sum_{k=1}^N (-1)^{k+1} \hat{\sigma}_k^z
\end{align} 
where $\hat{\sigma}^q_k (q\!\!=\!\!x,\!y,\!z)$ are the Pauli matrices on site $k$, \(J_k\) is the site-dependent coupling strength between the nearest-neighbor interacting spins and \((h_1^\prime, h_2^\prime)\)-pair represents the components of the uniform and alternating transverse magnetic fields. We consider two cases of \(\hat{\mathcal{H}}_1\), $XX$ model, where \(J_k=J\) i.e., site-independent coupling strength and the Su–Schrieffer–Heeger  model, where the interaction is pair-dependent. 

{\it Emergence of nH model in presence of bath.}
The imaginary alternating transverse magnetic field can be implemented via interaction with a monitored auxiliary qubits, where the effective dynamics is given by $\hat{\mathcal{H}}_{nH}$ with the post-selection in the no-click limit, as shown in Appendix~\ref{app:nH_deriv}. Specifically, each \(N\) sites interact locally with \(N\) auxiliary bath qubits on which projective measurements in \{\(|0\rangle,|1\rangle\)\} basis are performed continuously and the post-selection with \(|0\rangle\) outcome results in the nH model (see Fig. \ref{fig:sch}). Hence, considering state transfer in the presence of bath becomes extremely crucial.

{\it Average fidelity for $U(1)$-symmetric nH Hamiltonian.}
For the $U(1)$-symmetric Hamiltonian, denoting $\ket{\textbf{A}}\!\equiv\!\ket{\mathbf{1}}$ and $\ket{\textbf{B}}\!\equiv\!\ket{\mathbf{N}}$ for the sender's and the receiver's states respectively, the Haar averaged fidelity $\mathcal{F}_N(t)$ can be computed analytically~\cite{Bose2003},
\begin{align}
\nonumber \mathcal{F}_N(t) &= \frac{1}{2} + \frac{(|b_B|\cos\gamma - 2|b_B|^2 + \Gamma|b_B|\cos\gamma)}{(\Gamma-1)^2} \\ & + \frac{((1+\Gamma)|b_B|^2 - 2\Gamma|b_B|\cos\gamma) \log \Gamma}{(\Gamma-1)^3}
\label{eq:fid_eqn}
\end{align}
for the non-Hermitian Hamiltonian $\hat{\mathcal{H}}_{nH}$, with \(b_k \!=\! \langle \mathbf{k}|e^{-i \hat{\mathcal{H}t}}|\mathbf{A}\rangle\), $\Gamma \!=\! \sum_{k=1}^N b_k$ and \(|b_B|\cos\gamma\) is the real part of \(b_B\) (see Appendix~\ref{app:fid_deriv}). Note that in the Hermitian case, $\Gamma\!=\!1$, $\text{lim}_{\Gamma\to1}\mathcal{F}_N(t)$ is well-defined and gives the fidelity for QST with $U(1)$-symmetric Hermitian Hamiltonian.

\section{Non-hermitian \(\hat{\mathcal{P}}\hat{\mathcal{T}}\) symmetric $XX$ model Enables Higher-Fidelity Quantum State Transfer}
\label{sec:xx_model}

Let us first consider the $XX$ model with alternating field at neighboring sites, having \(J_k\!=\!J\ \forall k\!=\!1,\dots N\!-\!1\), i.e., uniform interaction between all the nearest-neighbor spin pairs, with $h_1\equiv h_1^\prime/J$ and $h_2\equiv h_2^\prime/J$. Hence, $\hat{\mathcal{H}}_1^{XX}$ is both $\hat{\mathcal{P}}$- and $\hat{\mathcal{T}}$-symmetric, whereas $\hat{\mathcal{H}}_2^{XX}$ breaks the $\hat{\mathcal{P}}$-symmetry. The non-Hermitian Hamiltonian $\hat{\mathcal{H}}_{nH}^{XX}\!=\!\hat{\mathcal{H}}^{XX}_1+i\hat{\mathcal{H}}_2^{XX}$ retains the overall \(\hat{\mathcal{P}}\hat{\mathcal{T}}\) symmetry (even \(N\)) and the nH $XX$ model has two regimes of spectrum: (1) unbroken, where all the eigenvalues are real and (2) broken, where at least one pair of complex conjugate eigenvalues emerge, depending upon the (\(h_1, h_2\)) values and the system size \(N\).

We aim to compare the fidelities of \(\hat{\mathcal{H}}_h\) and \(\hat{\mathcal{H}}_{nH}\) in the broken and unbroken regimes, with and without the presence of \(\hat{\mathcal{P}}\hat{\mathcal{T}}\) symmetry and identify the suitable parameter regimes where there can be potential benefit of non-Hermiticity in beating the classical limit or over Hermitian case.

\subsection{Advantage in pseudo-Hermitian model} 

Let us focus on the unbroken regime of the non-Hermitian \(\hat{\mathcal{P}}\hat{\mathcal{T}}\)-symmetric Hamiltonian, which exhibits only real eigenvalues. For a fixed \(N\), the fidelity of an arbitrary quantum state, computed using Eq. (\ref{eq:fid_eqn}), follows a similar profile with time corresponding to Hermitian case. However, such behavior crucially depends on \((h_1,h_2)\)-pair. In particular, there exist parameter regimes in which the trend of fidelity obtained from nH evolution at large time is different than that found through Hermitian Hamiltonian (see Figs. \ref{fig:plt_xx_ub} (a) and (b)). 
Extensive search in the \((h_1,h_2)\)-plane reveals some generic advantages of non-Hermiticity over the corresponding Hermitian model in the state transmission scheme.

{\it Observation 1: Advantage in minimum time. } It is possible to tune magnetic field strength in such a manner that the minimum time required to achieve fidelity above the classical threshold of \(2/3\) for \(\hat{\mathcal{H}}_{nH}^{XX}\) is lower than that can be obtained through \(\hat{\mathcal{H}}_{H}^{XX}\), i.e., 
\begin{equation}
\nonumber t_{nH}^{min} < t_{H}^{min},
\end{equation}
where \(t_x^{min}\) denotes the minimum time for which \(\mathcal{F}_N > 2/3\). For example, we find that for \(N=16\) and \((h_1=0.20,h_2=0.06)\), \(t_{nH}^{min} \approx43/J < t_H^{min}\) (see Figs \ref{fig:plt_xx_ub}(a) and (b) for more \((h_1,h_2)\)-pairs).

{\it Observation 2: Higher fidelity for nH model. } In the state transfer protocol, there are two important components - (1) the first maximum achievable fidelity, i.e., in the time profile of fidelity, when the fidelity reaches maximum value for the first time, denoted as  \(\mathcal{F}_N^{(1)}\), for a fixed \(N\), beyond \(\mathcal{F}_c\); (2) the minimum time (considered before) at which \(\mathcal{F}_N>2/3\). It is interesting to observe that the open system dynamics leading to non-Hermitian system can provide benefits with respect to the components (1) and (2). Specifically, we notice that 
\(\mathcal{F}^{(1)}_{N}\) corresponding to the non-Hermitian time-evolution can rise higher than that for the Hermitian system, i.e., 
\begin{equation}
\nonumber \mathcal{F}^{(1)}_{N,nH} > \mathcal{F}^{(1)}_{N,H},
\end{equation}
when \(N\) is reasonably moderate to high\footnote{Note that there is a cut-off in the system-size beyond which no choice of parameters can outperform the classical fidelity for both the Hermitian and non-Hermitian models. We later discuss the upper bound on \(N\) by optimizing over the parameter space of the driven Hamiltonian.}. For moderate system size, say \(N=16\), non-Hermitian dynamics can provide fidelity \(0.8\) with some \((h_1,h_2)\)-pairs while the evolution governed by the Hermitian $XX$ Hamiltonian can not surpass the classical fidelity (see Fig. \ref{fig:plt_xx_ub}(a)).


These observations lead to a natural question: {\it "Does any dedicated parameter regime exist where QST is benefited by the inclusion of non-Hermiticity in the evolution over its Hermitian analogue?"} In order to understand about the parameter regimes, we compute the first maximum fidelity, \(\mathcal{F}_N^{(1)}\), by varying the uniform and alternating magnetic fields, \(h_1 \in [0,0.5]\) and \(h_2 \in [0,0.3]\) in the  evolving Hamiltonian \(\hat{\mathcal{H}}_{nH}^{XX}\) and \(\hat{\mathcal{H}}_{H}^{XX}\). In the case of \(\hat{\mathcal{H}}_{nH}^{XX}\) with \(N=16\), we observe that the eigenspectrum is broken when \(h_2 \gtrsim 0.18\) (computed numerically for all values of \(0 \leq h_1 \leq 0.5\)) and it is separated from the unbroken regime, as depicted in Fig. \ref{fig:plt_xx_ub}(c).

In the unbroken regime, the entire set of parameter pairs $(h_1,h_2)$, except some patches, gives the fidelity $\mathcal{F}_N(t)>2/3$, for a particular local maxima. Interestingly, some ``bubble"-like patterns for \(\mathcal{F}_N^{(1)}\) emerge in the \((h_1,h_2)\)-plane, both for \(\hat{\mathcal{H}}_{nH}^{XX}\) and \(\hat{\mathcal{H}}_{H}^{XX}\). This is due to the choice of figure of merit \(\mathcal{F}_N^{(1)} >2/3 \) which cares only the {\it first local} maxima and classical limit. 
Hence, the bubble boundaries trace the time-evolving $\mathcal{F}_N^{(1)}$ for different (\(h_1,h_2\))-pairs. Such a behavior is universal, occurring in other $U(1)$-symmetric model examined here, i.e.,  in both Hermitian Hamiltonians and the unbroken phases of their non-Hermitian counterparts as will be seen also in the next section for the SSH model.
Note that when \(h_1=0\), the fidelity corresponding to both H and nH models oscillate about \(0.5\) and do not ever cross $\mathcal{F}_c$, signifying the necessity of transverse magnetic field component \(h_1\) in the state transfer protocol. Interestingly, near the exceptional line, the behavior is sensitive to the real uniform magnetic field $h_1$, and gives the highest value of \(\mathcal{F}_N^{(1)}\) for the nH model. Eg. with \(N=16\),  \(\max_{(h_1, h_2)}\mathcal{F}^{(1)}_{N, nH}\sim0.94\), whereas in the Hermitian case, \(\max_{(h_1, h_2)}\mathcal{F}^{(1)}_{N, H}\sim0.89\), highlighting the significance of the non-Hermitian framework.

\subsection{No-gain parameter region: Hermitian and non-Hermitian correspondence}
\label{subsec:herminonHcorr}

\begin{figure}
    \centering
\includegraphics[width=0.8\linewidth]{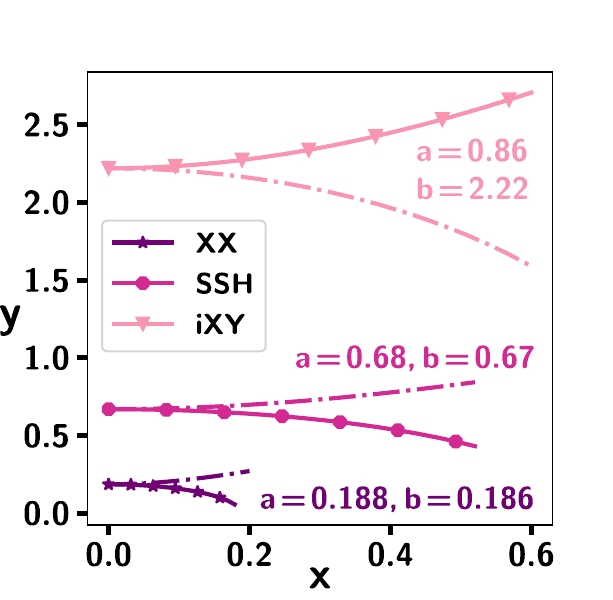}
    \caption{ {\bf Connecting parameter space of the pseudo non-Hermitian model  with the Hermitian ones.}  The lines represent the relation between \(y\equiv h_1\) (\(h\)) and \(x \equiv h_2\) (\(\gamma\)) (see Eqs. (\ref{eq:ellipse}) and (\ref{eq:hyperbola})) where the fidelity \(\mathcal{F}_N(t)\) could not reach the classical threshold of \(2/3\) for the $XX$, and SSH  (\(iXY\)) models.  The gradient of colors, from dark to light (pink), represent parameters of Eq. (\ref{eq:ellipse}) and (\ref{eq:hyperbola}) for the \(XX\), SSH and \(iXY\) models, with system-size $N=16$ for the $XX$ and the SSH model, while $N=8$ for the $iXY$ model. Solid lines and dashed-dot lines correspond to the nH and corresponding Hermitian models respectively. All the axes are dimensionless. 
    }
    \label{fig:corr}
\end{figure}

\begin{figure}
    \centering
\includegraphics[width=0.9\linewidth]{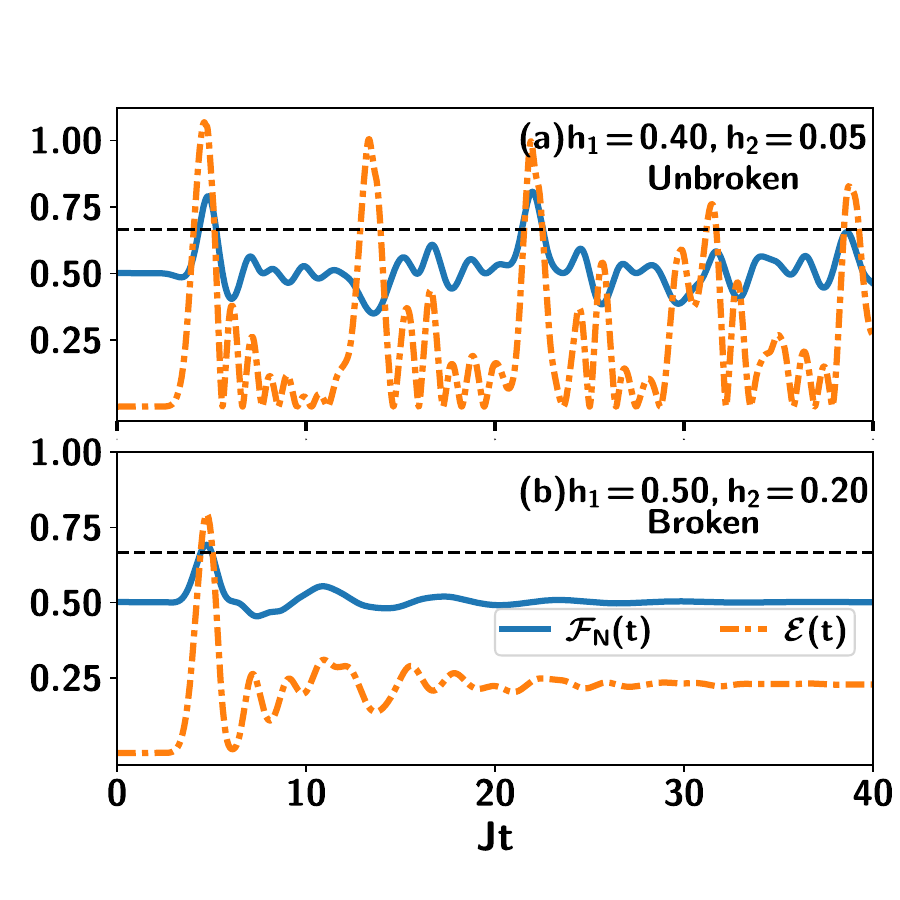}
    \caption{{\bf Profile of entanglement \(\mathcal{E}(t)\) (orange colored dashed-dot lines), and the corresponding fidelity \(\mathcal{F}_N(t)\) (solid blue lines), with respect to time \(t\).}  (a) Unbroken, and (b) broken regimes of the \(XX\) Hamiltonian. The maxima(s) of both \(\mathcal{F}_N(t)\) and \(\mathcal{E}(t)\) are aligned when \(\mathcal{F}_N (t) > \mathcal{F}_c\), thereby underscoring that the presence of entanglement is necessary for successful QST. Here, \(N=16\) and all the axes are dimensionless.
    }
    \label{fig:entvsfidelity}
\end{figure}

Our analysis reveals that there exist pairs of magnetic field strengths, \((h_1,h_2)\) that satisfy a specific relation for which the classical threshold cannot be surpassed, both in the non-Hermitian model within unbroken regime and in the corresponding Hermitian case. Remarkably, the parameter relation that fails to yield a quantum advantage in the unbroken non-Hermitian regime is directly connected to its Hermitian counterpart. In particular, for the pseudo-Hermitian case, the admissible parameter points lie on an ellipse described by 
\begin{eqnarray}
\frac{h_1^2}{b^2} + \frac{h_2^2}{a^2} = 1,
\label{eq:ellipse}
\end{eqnarray}
where \(a\) and \(b\) depend on the Hamiltonian, as shown in Fig.~\ref{fig:corr}. In contrast, in the Hermitian case, an analogous relation emerges upon replacing \(h_2\) by \(ih_2\), leading to an equation of hyperbola,
\begin{eqnarray}
 \frac{h_1^2}{b^2} + \frac{{(ih_2)}^2}{a^2} = 1 \rightarrow \frac{h_1^2}{b^2} - \frac{h_2^2}{a^2} = 1.
 \label{eq:hyperbola}
\end{eqnarray}
See Figs. \ref{fig:plt_xx_ub}(c) and (d) and Fig. \ref{fig:corr}. This behavior is found to be independent of moderate system size, \(N \leq 150\).
This shows that the Hermitian and non-Hermitian models in the unbroken regimes are the counterparts of each other and share a correspondence. Note that in a different spirit, the factorization surface in the Hermitian model and the exceptional surface in the corresponding non-Hermitian \(\hat{\mathcal{R}}\hat{\mathcal{T}}\)-symmetric model was shown to be connected~\cite{gan_adi_factor}. Further, Hermitian-non-Hermitian correspondence is observed also in the SSH  and \(iXY\) models, as discussed in Secs. \ref{sec:ssh_model} and \ref{sec:ixy_model}.
Therefore, it is tempting to conjecture that a generic relation, \(X_A\) (\(A = nH, H\)), in terms of parameters, \(x_1, x_2, x_3, \ldots\),  between Hermitian and non-Hermitian dynamics exists which highlights the condition for not obtaining nonclassical fidelity in the state transfer scheme,  i.e., 
\begin{eqnarray}
X_{nH}^{x_1,x_2,x_3,....} \leftrightarrow X_H^{ix_1,ix_2,ix_3,....},
\label{eq:correspondence}
\end{eqnarray}
where parameters, \(x_1, x_2, x_3, \ldots\) in the nH systems have to be replaced by \(ix_1, ix_2, ix_3, \ldots\). Note further that no nonclassical fidelity in the QST protocol in a system does not imply that the system does not possess any nonclassical feature like entanglement \cite{Horodecki2009_rmp} (see Fig. \ref{fig:entvsfidelity} in this context).

\subsection{Absence of \(\hat{\mathcal{P}}\hat{\mathcal{T}}\)-symmetry - Favorable for state transfer}

In the broken regime, the nH model possess eigenvectors, which break the \(\hat{\mathcal{P}}\hat{\mathcal{T}}\) symmetry of the Hamiltonian, with conjugate pairs of complex eigenvalues. Therefore, the time-evolution operator responsible for state to be transferred is \(e^{-i\hat{\mathcal{H}}t} = Pe^{-iEt}P^{-1}\), with \(E=\{\tilde{e}_k\}\) being the diagonal matrix of the eigenvalues and the matrix \(P\) incorporates the corresponding eigenvectors. Since few of the eigenvalues of \(\hat{\mathcal{H}}_{nH}\) are imaginary in the broken regime, this results in the time-evolution operator as \(e^{\pm |\tilde{e}_k|t}\). This leads to an exponential increase in the normalization factor $\Gamma$ in Eq.~(\ref{eq:fid_eqn}), essential for the non-Hermitian evolution. Denoting $\tilde{e}_{\max}$ as the imaginary eigenvalue with largest magnitude, the exponentially increasing \(e^{|\tilde{e}_{\max}|t}\) factor results in the evolution in the restricted vector space of eigenvectors with eigenvalues with imaginary part as $\tilde{e}_{\max}$. In the $U(1)$-symmetric case, the restricted subspace is defined by a unique eigenvector, and the fidelity saturates to \(0.5\) at longer times. Interestingly, at short times, the fidelity $\mathcal{F}_{N,nH}(t)$ oscillates and can also reach beyond $\mathcal{F}_c$ for certain $(h_1,h_2)$ parameter points in the broken regime due to the competition between the normalization and the overlap of the resulting state at site \(N\) with the desired input state, depicted in Fig.~\ref{fig:plt_xx_b}(a). In Appendix~\ref{sec:appen_xx}, the analytics for \(N=2\) clearly shows such trade-off. However, in this domain, we find the following:

{\it Observation 3: Hermitian model wins.} The first fidelity peak $\mathcal{F}_{N,nH}$ can exceed the classical threshold for some parameters, although it always remains below the fidelity attained by the Hermitian counterpart in the $\hat{\mathcal{P}}\hat{\mathcal{T}}$-symmetric $XX$ model.

{\it Absence of \(\hat{\mathcal{P}}\hat{\mathcal{T}}\) symmetry:} This obstacle can be overcome when the eigenspectrum of the non-Hermitian $XX$ model is always broken for odd number of sites and the model does not possess the \(\hat{\mathcal{P}}\hat{\mathcal{T}}\)-symmetry. In this situation, the fidelity profile with time displays a generic feature. Specifically, it shows an initial rise leading to $\mathcal{F}_{N,nH}^{(1)}$ followed by oscillations and a long-time decay towards \(0.5\) (see Fig.~\ref{fig:plt_xx_b}(a)). However, a crucial difference between odd and even sites in the broken regime emerges.

{\it Observation 4: Advantage via non-\(\hat{\mathcal{P}}\hat{\mathcal{T}}\)-symmetric model.} The non-Hermitian system can attain fidelities that not only exceed the classical \(2/3\) limit but also surpass those of the Hermitian case when the system does not possesses \(\hat{\mathcal{P}}\hat{\mathcal{T}}\) symmetry (see Fig.~\ref{fig:plt_xx_b}(b)).


\begin{figure}
    \centering
\includegraphics[width=0.9\linewidth]{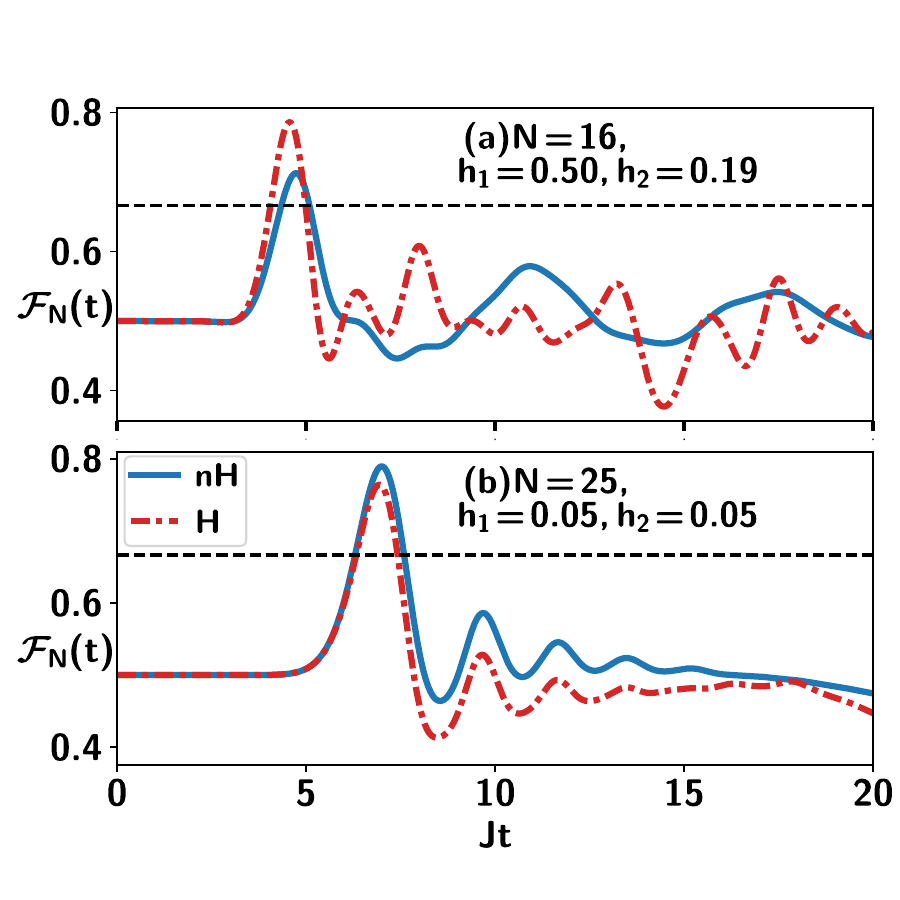}
    \caption{Illustration of fidelity \(\mathcal{F}_N(t)\) with respect to time is presented for (a) \(N=16, h_1 = 0.5, h_2 = 0.19\), and (b) \(N=25, h_1 = 0.05, h_2 = 0.05\). Both these parameter sets considered belong to the broken regime of the  non-Hermitian \(XX\) Hamiltonian. (a) The first maxima of fidelity obtained via non-Hermitian evolving Hamiltonian with \(\mathcal{\hat{P}\hat{T}}\)-symmetry  rises above the classical limit of \(2/3\) but remains below the   fidelity through the corresponding Hermitian model. In contrast, in (b), it beats the performance via the Hermitian model, thereby establishing the advantage of non-Hermiticity. Note that the nH \(XX\) model in this case lacks \(\mathcal{\hat{P}\hat{T}}\)-symmetry and  the number of sites is odd. All the axes are dimensionless.}
    \label{fig:plt_xx_b}
\end{figure}

\subsection{Connection between distribution of entanglement  and fidelity profile}

For the quantum state transfer protocol, quantum correlations, especially entanglement, between the sender and the receiver are required to surpass the classical threshold $\mathcal{F}_c$. Therefore, the entanglement generation capability of the non-Hermitian Hamiltonian $\hat{\mathcal{H}}_{nH}$ plays a significant role to aid the success of QST and we analyze the entanglement profile of the system, by computing the entanglement of Bob's spin at site $N$ with all the other \((N-1)\) spins, including Alice's, with the variation of  time. Specifically, using logarithmic negativity \cite{vidal_pra_2002, plenio_prl} as the entanglement measure $\tilde{\mathcal{E}}$, we compute 
\begin{equation}
    \nonumber \mathcal{E}(t) \!=\!\!\sum\limits_{k=1}^{N-1}\tilde{\mathcal{E}}(\rho_{(k,N)}(t));\quad \tilde{\mathcal{E}}(\rho_{(k_1,k_2)}) \!=\! \log_2(2\mathcal{N}_{(k_1,k_2)}+1),
\end{equation}
where $\rho_{(k_1,k_2)}(t)$ is the time-evolved reduced density matrix of two sites,  $k_1$ and $k_2$, and $\mathcal{N}_{(k_1,k_2)}$ represents the negativity, defined as the absolute sum of negative eigenvalues of the partially transposed state, $\rho_{(k_1,k_2)}$$\rho^{T_{k_1}}_{(k_1,k_2)}$ \cite{peres_prl_1996, horodecki_pla_1996}. Starting from the product state, the dynamics via interacting Hamiltonian leads to a generation of entanglement among different two parties, \(\rho_{k,N}(t)\) (\(k=1, 2, \ldots, N-1\)),  thereby creating a first peak in \(\mathcal{E}_N(t)\), followed by its 
oscillatory behavior with time.  We indeed find a connection between the peaks of fidelity $\mathcal{F}_N(t)$ and that of entanglement $\mathcal{E}(t)$. In particular,  when the state transfer protocol is carried out with the nH \(XX\) model,  $\mathcal{E}(t)$ is locally maximum at all peaks of $\mathcal{F}_N(t)>\mathcal{F}_c$, although we also observe situations in which $\mathcal{F}_N(t)<\mathcal{F}_c$ even when $\mathcal{E}(t)$ reaches local maximum. This observation is found to be true for both unbroken and broken regimes (as shown in Fig. \ref{fig:entvsfidelity}), and both nH and H models. This investigation can be summarized as follows:

{\it Observation 5: Sufficient amount of entanglement is necessary for state transfer.} For the fidelity to exceed the classical limit, the presence of entanglement is necessary but not sufficient. While the local maxima of both quantities nearly coincide in time when the fidelity crosses the \(\mathcal{F}_c\)
threshold, the reverse implication does not always hold.

\begin{figure*}
    \centering
\includegraphics[width=\linewidth]{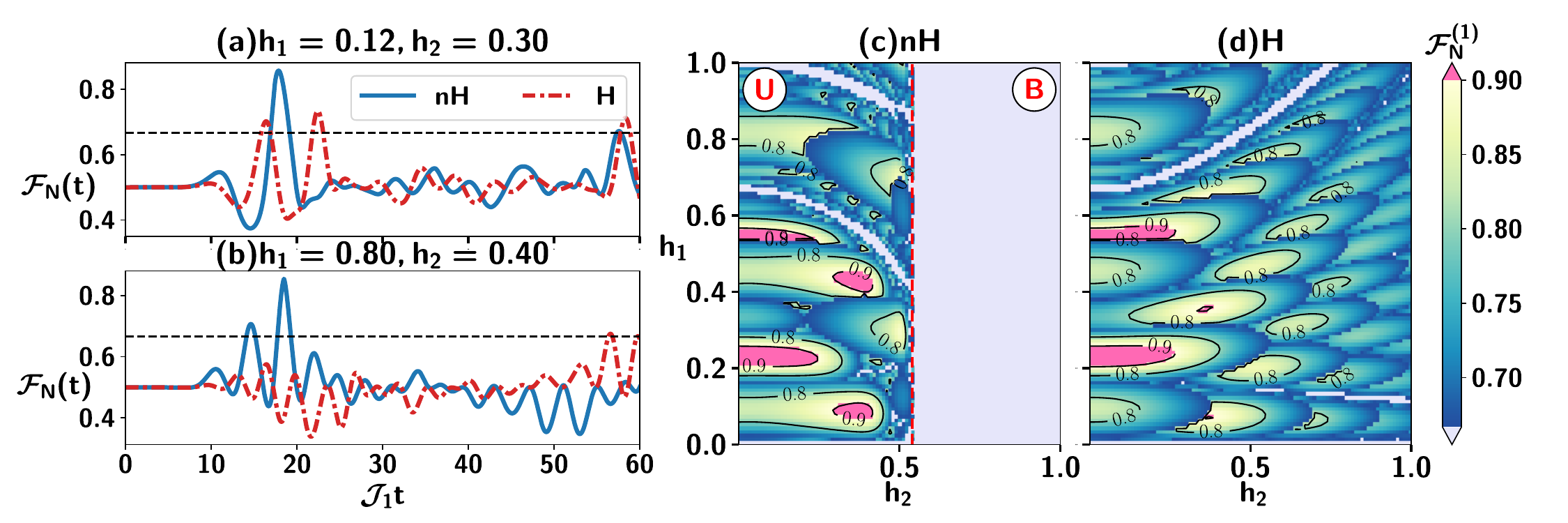}
    \caption{ \(\mathcal{F}_N(t)\) (ordinate) vs time \(t\) (abscissa) for  (a) \(h_1 = 0.12, h_2 = 0.30\) and (b) \(h_1 = 0.80, h_2 = 0.40\), considering SSH model with \(\mathcal{J}_2 = 0.5<1\), with both parameter regimes in the unbroken regime. 
    Note that the nH model leads to a higher first maximum value of fidelity $\mathcal{F}_N^{(1)}$  as compared to the Hermitian counterpart. (c) and (d) represent the contour plots of nH- and H- SSH models respectively, with respect to  (\(h_2,h_1\))-pair and $\mathcal{F}_N^{(1)}$ as the $z$-axis. The (red) dashed line, i.e., \(h_2 \lesssim 0.5\), separates the broken (B) regime from the unbroken (U) one in (c), the parameter points with $\mathcal{F}_N(t)<\mathcal{F}_c$ are marked lavender and follow Eq.~ (\ref{eq:ellipse}) and Eq.~ (\ref{eq:hyperbola}) for (c) and (d) respectively. 
    The parameters \((a,b)\) in this case read (\(0.68,0.67\)). . The pink colored points represent \(\mathcal{F}_N^1>0.9\). Here \(N=16\).
    All the axes are dimensionless. }
        \label{fig:plt_ssh_0.5}
\end{figure*}

\section{Non-hermitian \(\mathcal{\hat{P}\hat{T}}\)- symmetric SSH model}
\label{sec:ssh_model}

Let us now explore whether the  benefit of QST through the non-Hermitian $XX$ Hamiltonian over its Hermitian counterpart is a generic characteristic or a system-specific feature. To address this question, let us consider the SSH model, with non-uniform interaction strength, while preserving the $U(1)$- and \(\hat{\mathcal{P}}\hat{\mathcal{T}}\)-symmetry of the $XX$ model. The corresponding Hamiltonian for even \(N=2n\) sites with \(n\) cells (a cell is pair of two sites $(2j-1, 2j)$ for $j=1,2,\dots, n$) can be written as~\cite{ssh_og,Asboth2016}
\begin{align}
\nonumber \hat{\mathcal{H}}_1 &= \mathcal{J}_1^{\prime}\sum_{k\in odd} (\sigma^x_k\sigma^x_{k+1} + \sigma^y_k\sigma^y_{k+1}) \\  &+\mathcal{J}_2^{\prime}\sum_{k\in even} (\sigma^x_k\sigma^x_{k+1} + \sigma^y_k\sigma^y_{k+1}) - h_1^{\prime}\sum_{k=1}^N \sigma_k^z,
\end{align} 
where \(J_k = \mathcal{J}_1^{\prime}\) for \(k=1,3,\dots 2n-1\) i.e., the intra-cell hopping amplitude, while,  \(J_k = \mathcal{J}_2^{\prime}\) for \(k=2,4,\dots 2n\), representing the inter-cell hopping amplitudes, 
and \(\left(\frac{\mathcal{J}_1^{\prime}}{\mathcal{J}_1^{\prime}}\!=\!1, \frac{\mathcal{J}_2^{\prime}}{\mathcal{J}_1^{\prime}}\equiv \mathcal{J}_2\right)\) are the parameters of the model with the dimensionless magnetic field strengths $h_1\equiv h_1^\prime/\mathcal{J}_1^{\prime}$ and $h_2\equiv h_2^\prime/\mathcal{J}_1^{\prime}$. The Hermitian SSH model, $\hat{\mathcal{H}}_1$, without the magnetic field $h_1\!=\!0$  undergoes a quantum phase transition at \(\mathcal{J}_2 = 1\), which corresponds to the $XX$ model~\cite{ssh_og,Asboth2016}. It was shown that its non-Hermitian  counterpart obtained by adding imaginary on-site potential also has a rich phase diagram  due to the competition between \(\mathcal{J}_1^{\prime}\) and \(\mathcal{J}_2^{\prime}\)~\cite{Halder_2023}. In our study, we investigate the behavior of fidelity in the two regimes of the non-Hermitian model \(\hat{\mathcal{H}}_1 + i \hat{\mathcal{H}}_2\) having imaginary alternating magnetic field as before -- (R1) \(\mathcal{J}_2 < 1\), i.e., intra-cell interactions are stronger than the intra-cell ones, and (R2) \(\mathcal{J}_2 > 1\). 

\subsection{\(\mathcal{J}_2<1\) -- Beneficial unbroken vs no gain in broken phase  }

In the regime \(\mathcal{J}_2 < 1\), the variation of magnetic field strength \((h_1, h_2)\)-duo again leads to the transition from unbroken to broken regimes. In the context of state transfer,  we find that the power of the evolution operator in the unbroken phase  differs from the broken one.


{\it Unbroken regime.} Observations 1 and 2  also hold for this model  as seen in Fig. \ref{fig:plt_ssh_0.5} (a) and (b). In contrast to the $XX$ model, there are larger number of parameter sets in the nH case as compared to the Hermitian counterpart, where the fidelity reaches as high as \(0.9\) and beyond, as shown by pink colored areas in the Fig. \ref{fig:plt_ssh_0.5} for $N=16$. Note, however, that the highest values obtained through the non-Hermitian and the Hermitian model
remains same upto the numerical accuracy. Furthermore, in this model also, we observe that (a) the first maxima of fidelity  follows bubble structure, both in H and nH models; (b)   the relation between parameters in the nH  and the H models in which the fidelity cannot reach the classical threshold again follow the equations of ellipse and hyperbola respectively, as conjectures in Sec. \ref{subsec:herminonHcorr}.  


\begin{figure*}
    \centering
\includegraphics[width=\linewidth]{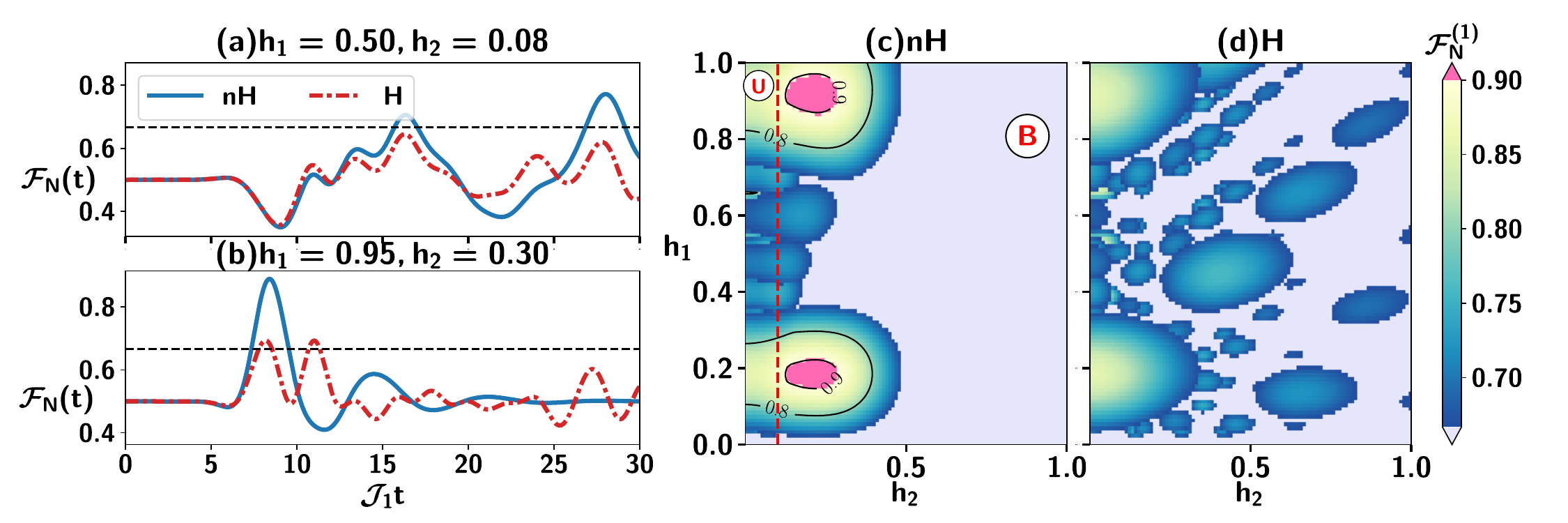}
    \caption{ {\bf Non-Hermitian evolution wins.}  \(\mathcal{F}_N(t)\) (ordinate) with time \(t\) (abscissa) for (a) \(h_1 = 0.50, h_2 = 0.08\) from the unbroken regime, and (b) \(h_1 = 0.95, h_2 = 0.30\) from the broken regime, which is represented by the dotted red vertical line in (c). Unlike Fig. \ref{fig:plt_ssh_0.5},   \(\mathcal{J}_2 = 1.2 >1\) in this case. The contour plots are presented for the analysis of parameter regimes where QST advantage occurs in (a) non-Hermitian and (b) Hermitian models.  Note that fidelity in the broken regime can reach \(0.9\) which cannot be obtained by Hermitian evolution. Clearly, in all the plots, non-Hermitian dynamics, especially in the broken regime, outperforms the Hermitian ones. All the axes are dimensionless. }
    \label{fig:plt_ssh_1.2}
\end{figure*}

{\it No quantum advantage in broken regime.} Within the broken regime, no choice of parameters yields a fidelity that surpasses the classical threshold, \(\mathcal{F}_c =2/3\). This observation is independent of system-size \(N\leq 150\).  Interestingly, the non-beneficial region spanning the entire broken phase is a specialty of the SSH model with \(\mathcal{J}_2 < 1\) which is not true when \(\mathcal{J}_2 = 1\) (the \(XX\) model) as well as \(\mathcal{J}_2 > 1\) which will be considered next. 


\subsection{\(\mathcal{J}_2>1\): Broken phase with non-Hermitian benefit over Hermitian model}

When \(\mathcal{J}_2>1\), the eigenspectrum is  broken for smaller values of $h_2$ (eg.  \(h_2 \gtrapprox 0.1\) for \(\mathcal{J}_2 = 1.2\) and \(N=16\)). The value of \(h_2\), above which the spectrum gets imaginary, decreases with the increase of \(\mathcal{J}_2\), implying that the spectrum gets completely broken for higher inter-cell hopping strengths. Eg.   with \(\mathcal{J}_2 = 2.0\), \(h_2 = 0.01\) above which the system is broken for all values of \(h_1\). 

{\it Broken region is special.} First, we observe that \(\mathcal{F}^{(1)}_{nH}\) increases with the increase of \(\mathcal{J}_2\). For instance, our numerical simulation for \(N=16\) reveals that  the fidelity \(\mathcal{F}^{(1)}_{16,nH}= 0.91\) for \(\mathcal{J}_2 = 1.2\) increases  to \(\mathcal{F}^{(1)}_{16,nH}=0.95\) for \(\mathcal{J}_2 = 2\). Interestingly, the corresponding Hermitian case shows the opposite behavior -- it yields \(0.86 \) to \(\mathcal{F}_c\) by varying  \(\mathcal{J}_2\) from  \(1.2\) to \(2\), thereby showing non-beneficial role of high value of \(\mathcal{J}_2\) in Hermitian SSH model. 
Therefore, this broken phase with dominant inter-cell hopping highlights the maximum benefit of non-Hermiticity in terms of the first maximum fidelity in the state transfer protocol as compared to the corresponding Hermitian model. It will be intriguing to study whether such a benefit persists even for other non-Hermitian models where the non-Hermiticity can be introduced in the system in a different way, like considering imaginary couplings which we will consider next.


\begin{figure}
    \centering
\includegraphics[width=0.9\linewidth]{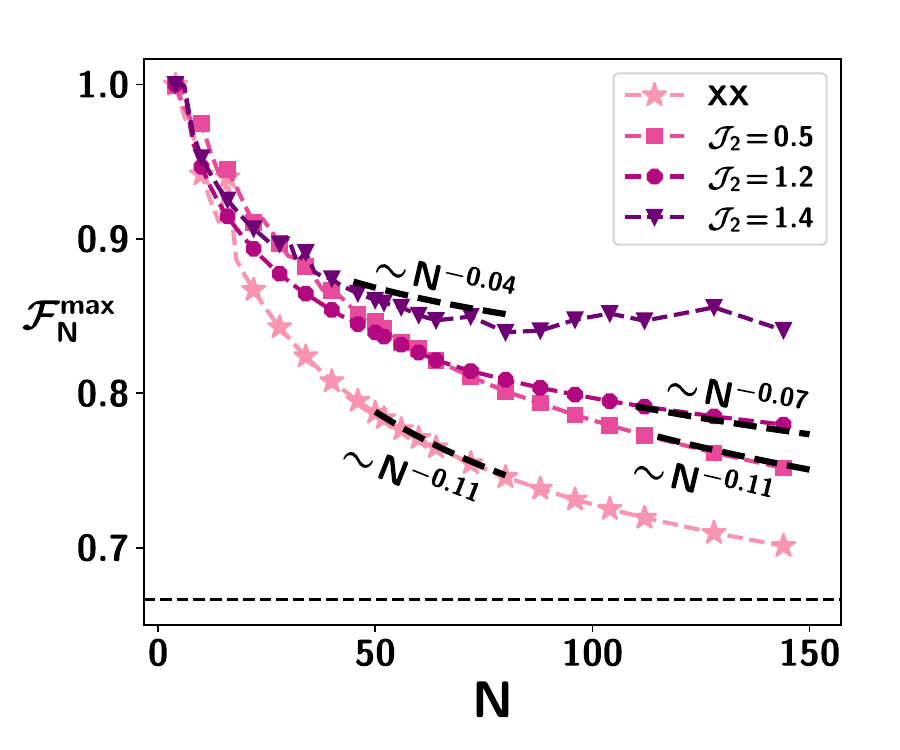}
    \caption{ Maximum fidelity optimized over \((h_1,h_2)\) values denoted as \(\mathcal{F}_N^{max}\) (ordinate) against the system size, \(N\) (abscissa) for different values of \(\mathcal{J}_2\). While \(\mathcal{F}_N^{max}\sim N^{-0.11}\) for QST with the $XX$ model and the SSH model when $\mathcal{J}_2<1$,  the dependence on $N$ decreases with increasing $\mathcal{J}_2$, indicating possible QST over large distance. All axes are dimensionless.
    }
    \label{fig:fidvsN}
\end{figure}
\section{System size dependence in QST}
\label{sec:systemsize}

In the QST protocol,  it is well established that beyond a certain system size the fidelity cannot surpass the classical limit and also it decreases as the length of the chain used as a channel increases. For example, the Heisenberg model exhibits a quantum advantage only up to spin chains of length \(N=80\) \cite{Bose2003} while in the  \(XY\) model, this benefit can persist for significantly longer chains, extending to \(N=240\) \cite{bayat_2011}.  It is, therefore, intriguing to examine, for non-Hermitian \(XX\) and SSH models, how the fidelity decays with system size and if the decay rate depends on the system being in the broken or unbroken phase.


In order to address this query, for a given \(N\), we compute the first maxima of the fidelity  after maximizing over the magnetic field strengths \((h_1, h_2) \in [0,1]\). Mathematically, it reads as \(\mathcal{F}^{\max}_N = \max_{h_1,h_2}\mathcal{F}_N^{(1)}\) for the \(XX\) model. We observe that \(\mathcal{F}^{\max}_N\) decreases with the increase of \(N\), specifically as $N^{-0.11}$ for the $XX$ model. 

By fixing \(\mathcal{J}_2\) for the SSH model,  we again optimize \(\mathcal{F}_N^{(1)}\) over the \((h_1, h_2)\) parameters in the case of the SSH model. As reported before, QST depends on the values of \(\mathcal{J}_2\)s, and hence we consider the situation when \(\mathcal{J}_2 < 1\) and \(\mathcal{J}_2 > 1\). When intra-cell interaction dominates \(\mathcal{J}_2<1\), similar scaling with $N$ is obtained, as seen in the $XX$ model, as shown in Fig.~\ref{fig:fidvsN}. Interestingly, on stronger inter-cell interactions, i.e., \(\mathcal{J}_2 > 1\), the advantage seen in the broken regime persists for larger values on system size $N$, with the scaling of $N^{-0.07}$ for $\mathcal{J}_2=1.2$ and $N^{-0.04}$ for $\mathcal{J}_2=1.4$, highlighting the advantage of stronger inter-cell hopping for larger system-size as well.

\section{non-hermitian \(\hat{\mathcal{R}}\hat{\mathcal{T}}\) symmetric $iXY$ model}
\label{sec:ixy_model}

\begin{figure}
    \centering
\includegraphics[width=1.0\linewidth]{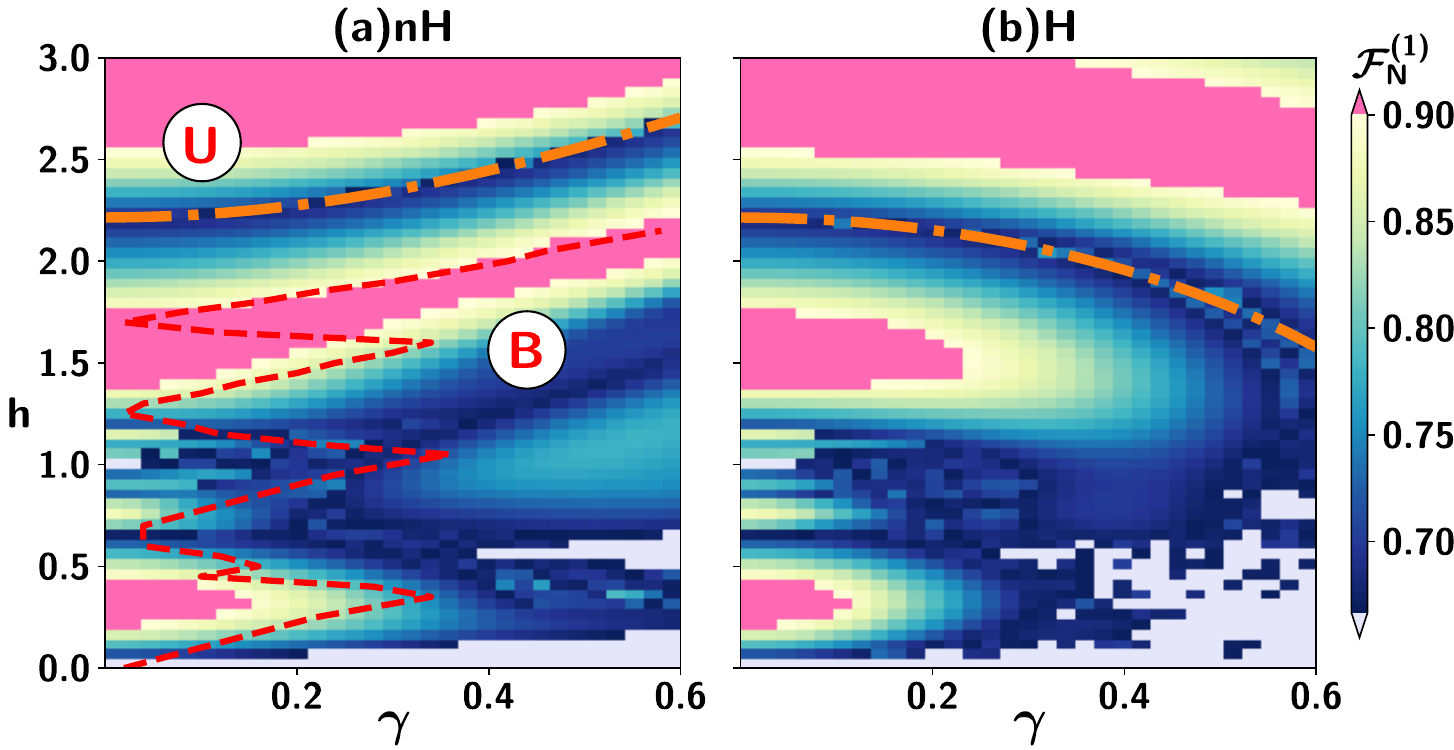}
    \caption{The contour plot of \(\mathcal{F}_N^{(1)}\) for (a) non-Hermitian \(\hat{\mathcal{R}}\hat{\mathcal{T}}\)-symmetric \(iXY\) model, and (b) the corresponding Hermitian \(XY\) model. The dashed red line distinguishes the unbroken (U) from the broken (B) regime. The dash-dotted (orange) lines represent the \((\gamma,h)\)-pairs where \(\mathcal{F}_N^{(1)}\) just reaches \(\mathcal{F}_c\) with $h>1$, which correspond to a hyperbola in (a) and an ellipse in (b) with parameters \(a=0.86\) and \(b=2.22\). Again, we find the correspondence between the Hermitian and non-Hermitian models as discussed in Sec. \ref{subsec:herminonHcorr}, suggesting this connection to be generic in nature.   Here, \(N=8\), and all axes are dimensionless.}
    \label{fig:plt_xy}
\end{figure}

After demonstrating the benefit of \(\mathcal{\hat{P}\hat{T}}\)-symmetric non-Hermitian Hamiltonian in the quantum state transfer, let us now carry out the evolution  through the \(\mathcal{\hat{R}\hat{T}}\)-symmetric non-Hermitian model (Appendix \ref{sec:appen_def}), namely the \(iXY\) model~\cite{Song_RT_symm, gan_adi_factor, Agarwal2023May, kda2024}. Taking \(\hat{\mathcal{H}}_1\) and \(\hat{\mathcal{H}}_2\) in Eq. (\ref{eq:fid_eqn}) as
\begin{align}
 \hat{\mathcal{H}}_1 &= J\!\sum_{k=1}^{N-1} (\hat{\sigma}^x_k\hat{\sigma}^x_{k+1} \!+\! \hat{\sigma}^y_k\hat{\sigma}^y_{k+1}) + h^{\prime}\sum_{N=1}^N \hat{\sigma}_k^z, \\
 \hat{\mathcal{H}}_2 &= J\gamma\!\sum_{k=1}^{N-1}(\hat{\sigma}^x_k\hat{\sigma}^x_{k+1} \!-\! \hat{\sigma}^y_k\hat{\sigma}^y_{k+1}),
\end{align} 
where \(\gamma\) is the anisotropy parameter, \(h\equiv h^{\prime}/J\) is the uniform transverse magnetic field and \(J\) is the interaction strength,  we get $\hat{\mathcal{H}}_{nH}$ as the $iXY$ Hamiltonian with $\hat{\mathcal{H}}_{H}$ as the corresponding Hermitian $XY$ model. The non-Hermitian $iXY$ Hamiltonian can also be realized via reservoir engineering~\cite{Fazio_reservoir_engineering_2018, Agarwal2023May}, where two adjacent sites interact with a common bath, resulting in non-Hermiticity in the interaction term. Note that this model lacks $U(1)$-symmetry, and hence the derived expression of fidelity  does not hold and is thus calculated numerically via Haar average.

In the case of evolution by the $iXY$ model, we observe that the QST with high fidelity (\(\approx 0.9\)) can be obtained in two situations -- (1) When both  magnetic field strength $h$, and anisotropy $\gamma$, are weak and the system lies in the unbroken regime; (2) For large magnetic field $h$, the QST is successful for various $\gamma$ values (increasing with increasing $h$ although not ubiquitous), with maximum fidelity near the exceptional points as shown in Fig.~\ref{fig:plt_xy}(a). 
Further, with low values of $h$, fidelity decreases with the increase of the non-Hermiticity parameter $\gamma$. 

Notice that although there are regions where non-Hermitian Hamiltonian can only provide benefit, there are parameter space where Hermitian model is also capable to transfer quantum state with a high fidelity. 

We again find a connection between the parameters in the Hermitian and the corresponding non-Hermitian models as given in Eq.~(\ref{eq:correspondence}). However, in the \(iXY\) model, similar relation with the parameters $(\gamma,h)$ is obtained  when the first maximum of the fidelity $\mathcal{F}_N^{(1)}$ is close to the classical threshold of \(\mathcal{F}_N^{(1)}\gtrsim \mathcal{F}_c\).  Note here that due to the absence of \(U(1)\)-symmetry in this model, we cannot simulate the evolution for high system size and so the observed $\mathcal{F}_N^{(1)} \gtrsim 2/3$ can be due to the finite-size.  Specifically,  the \((\gamma, h)\)-pair (for \(h>1\)) follows a hyperbola equation in the non-Hermitian case, while it becomes the equation of the ellipse  in the Hermitian case, with same constants \((a,b)\) in Eq. (\ref{eq:ellipse}), as depicted in Fig.\ref{fig:plt_xy}(a) and (b) for system size $N=8$. Note that the fidelity does not reach the classical threshold of \(\mathcal{F}_c=2/3\) when \(h=0\), \(\forall \gamma\). This again signifies the importance of the external transverse magnetic field, with larger magnetic field required for larger values of $\gamma$, in order to achieve \(\mathcal{F}_N^{(1)}> \mathcal{F}_c\) both in the Hermitian and the non-Hermitian case.


\section{Conclusion}
\label{sec:conclu}

High-fidelity transmission of quantum states between distant parties is a fundamental requirement for quantum communication and a wide range of quantum information–processing tasks. Quantum state transfer (QST) offers a natural framework for this purpose, where an interacting Hamiltonian on a network enables the transmission of quantum information, with a spin chain effectively acting as a quantum data bus. Since perfectly isolated systems are idealized, realistic implementations inevitably involve environmental interactions, which can be effectively described by non-Hermitian dynamics, thereby opening new possibilities for quantum protocols operating through non–completely positive trace-preserving (non-CPTP) channels.

This work explored the role of non-Hermiticity in enabling quantum state transfer across a chain of interacting qubits coupled to a continuously monitored auxiliary. Focusing on models with 
\(U(1)\)-symmetry and initially polarized states, we derived a general expression for the state-transfer fidelity in non-Hermitian systems. We then analyzed $\hat{\mathcal{P}}\hat{\mathcal{T}}$-symmetric \(XX\) and SSH models, subjected to real uniform and imaginary alternating magnetic fields. For the $XX$ model, we found that non-Hermitian dynamics can yield high-fidelity state transfer in parameter regimes belonging to the unbroken phase, where the corresponding Hermitian model does not perform as well. However, the maximum achievable fidelity remains identical in both Hermitian and non-Hermitian cases, allowing reliable state transfer over spin chains of moderate to large system size.
We further uncovered a simple correspondence between parameter regimes in which both Hermitian and non-Hermitian models fail to surpass the classical fidelity threshold, a relation that holds for both the \(XX\) and SSH models when intra-cell couplings dominate. Strikingly, when inter-cell interactions dominate and the system lies in the broken phase, the non-Hermitian SSH model enables very high-fidelity state transfer for moderate system sizes, in stark contrast to its Hermitian counterpart, which fails under the same conditions. This clearly demonstrates the advantage of non-Hermitian dynamics for QST in the presence of decoherence. Extending our analysis to the $\hat{\mathcal{R}}\hat{\mathcal{T}}$-symmetric $iXY$ model, we observed similar behavior, including a Hermitian–non-Hermitian correspondence in the no-gain of state transmission limit.

Summarizing, our findings highlight that the resources required to establish a robust quantum link depend on a subtle interplay between several factors, including the entangling-disentangling capability of the evolving Hamiltonian, the distance between the sender and the receiver, and the operational timescale to obtain the desired state. In this respect, our results offer hope that efficient quantum state transfer can remain feasible even in noisy, realistic quantum devices.

\section{Acknowledgments}
We acknowledge support from the project entitled ``Technology Vertical - Quantum Communication'' under the National Quantum Mission of the Department of Science and Technology (DST)  (Sanction Order No. DST/QTC/NQM/QComm/2024/2 (G)). We acknowledge the use of \href{https://github.com/titaschanda/QIClib}{QIClib} -- a modern C++ library for general-purpose quantum information processing and quantum computing (\url{https://titaschanda.github.io/QIClib}) and high-performance computing facility at Harish-Chandra Research Institute. This research was supported in part by the INFOSYS scholarship for senior students.

\appendix

\section {Definitions related to symmetries}
\label{sec:appen_def}

The parity symmetry \(\hat{\mathcal{P}}\) corresponds to the reflection in the position space, viz., \(\vec{x} \rightarrow -\vec{x}\) and \(\vec{p} \rightarrow -\vec{p}\) with $\hat{\mathcal{P}}^{\dagger}\!=\!\hat{\mathcal{P}}$, and is implemented in the chain of $N$ qubits as $\hat{\mathcal{P}}\rho_k \hat{\mathcal{P}} \!=\! \rho_{N-k+1}$. This corresponds to the reflection at the center in a plane perpendicular to the chain. The rotation symmetry, $\hat{\mathcal{R}}\!=\! \bigotimes_{k=1}^{N}\exp\left(-i\frac{\pi}{4}\hat{\sigma}^z_k\right)$, is the rotation by $\frac{\pi}{4}$ angle around the $z$-axis, which results to $\hat{\mathcal{R}}\hat{\sigma}^\alpha_k\hat{\mathcal{R}}^{\dagger} = -\hat{\sigma}^\alpha_k$ for $\alpha=x,y$ and $\hat{\mathcal{R}}\hat{\sigma}_k\hat{\mathcal{R}}^{\dagger} = \hat{\sigma}^z_k$, where $\hat{\sigma}^\alpha_k(\alpha=x,y,z)$ denotes Pauli matrices for all sites $k$. 

The anti-unitary symmetry is the time reversal symmetry \(\hat{\mathcal{T}}\), given by \(t \rightarrow -t\), and is represented by the complex conjugation with \(i \rightarrow -i\), i.e., $\hat{\mathcal{T}}\hat{\mathcal{H}}\hat{\mathcal{T}}\!=\!\hat{\mathcal{H}}^{*}$, for an arbitrary $\hat{\mathcal{H}}$. A Hamiltonian $\hat{\mathcal{H}}$ is said to be $\hat{\mathcal{S}}\hat{\mathcal{T}}$-symmetric (with $\hat{\mathcal{S}}\equiv\hat{\mathcal{P}},\hat{\mathcal{R}}$), when $[\hat{\mathcal{H}}, \hat{\mathcal{S}}\hat{\mathcal{T}}]=0$, which implies $\hat{\mathcal{S}}^{\dagger}\hat{\mathcal{H}}\hat{\mathcal{S}}=\hat{\mathcal{H}}^{*}$.


\section{Effective non-Hermitian evolution}
\label{app:nH_deriv}
A system of $N$ sites, evolving with a Hermitian Hamiltonian $\hat{\mathcal{H}}_1$, undergoes unitary dynamics. A non-unitary dynamics occur when the sites are weakly measured by interaction with continuously monitored auxiliary qubits. A qubit on site $k$, in state $\rho_k$, interacts with an auxiliary qubit $k^\prime$, initialized in a pure state $\ket{\omega}\!\bra{\omega}_{k^\prime}$, with the joint state as
\begin{align}
    &\tilde{\rho}^{(g)}_{k k^\prime} \!= \hat{U}^{(g)}_{kk^\prime}\left[\rho_{k} \otimes \ket{\omega}\!\bra{\omega}_{k^\prime}\right] \hat{U}^{(g)}_{kk^\prime} \nonumber\\
    &\quad\ \ = \sum_{\mathclap{m_1,m_2=0,1}} K_{m_1,k}^{(g,\omega)}\rho_k K_{m_2,k}^{(g,\omega)\dagger} \otimes \ket{m_1}\!\bra{m_2}_{k^\prime}, \nonumber\\
    &\hat{U}^{(g)}_{kk^\prime} \!= \exp[-i\epsilon (\hat{n}^{(g)}_k \!\!\otimes\! \hat{\sigma}^x_{k^\prime})];\quad \hat{n}^{(g)}_k \!=\! \frac{\hat{1}_k \!+\! (-1)^g\hat{\sigma}^z_k}{2}, \\
    &K_{m,k}^{(g,\omega)} \!=\! \sum_{\mathclap{a_1,a_2=0,1}} \left[ (\bra{a_1}_k \!\otimes\! \bra{m}_{k^\prime}) \hat{U}^{(g)}_{kk^\prime} (\ket{a_2}_{k} \!\otimes\! \ket{\omega}_{k^\prime}) \right] \ket{a_1}\!\bra{a_2}_k,\nonumber
\end{align}
where $\hat{n}^{(g)}_k \!=\! \ket{g}\!\bra{g}_k (\!g\!=\!0,\!1\!)$ are the projectors onto the eigenstates on $\hat{\sigma}^z_k$, giving two different unitary interactions, which leads to two possible sets of site dependent Kraus operators $\mathcal{K}_k \!=\! \{K_{m,k}^{(g,\omega)}\}_m$, for a given choice of $\ket{\omega}_{k^\prime}$. We choose $g\!=\!0$ for odd and $g\!=\!1$ for even sites $k$, and correspondingly the auxiliary is initialized in $\ket{\omega}_{k^\prime} \!=\! \ket{g}_{k^\prime}$ to obtain a $\hat{\mathcal{P}}\hat{\mathcal{T}}$-symmetric imaginary alternating field on a open chain of qubits. This gives the Kraus operators as
\begin{equation}
    \mathcal{K}_k \!=\!\! 
    \begin{cases}
        K_{0,k}^{(0,0)} \!\!=\! i\sin\epsilon\ket{0}\!\!\bra{0}, \ K_{1,k}^{(0,0)} \!\!=\! \cos\epsilon\ket{0}\!\!\bra{0} \!+\! \ket{1}\!\!\bra{1};
        \\
        K_{0,k}^{(1,1)} \!\!=\! \ket{0}\!\!\bra{0} \!+\! \cos\epsilon\ket{1}\!\!\bra{1}, \ K_{1,k}^{(1,1)} \!\!=\! i\sin\epsilon\ket{1}\!\!\bra{1},
    \end{cases}
    \label{eq:def_K}
\end{equation}
where $K_{m,k}^{(0,0)}$ are for odd sites $k$, and $K_{m,k}^{(1,1)}$ acts on even $k$ for $m\!=\!0,1$.

Therefore, measuring the auxiliary qubit $k^\prime$ on the eigenstate of $\hat{\sigma}^z$ results in the action of the outcome dependent Kraus operator on the site $k$. For a $N$-qubit state $\rho$, let us denote the state $\rho_k\!=\!\text{Tr}_{\bar{k}}(\rho)$, as the state at each site $k$, where $\text{Tr}_{\bar{k}}(\cdot)$ is the partial trace over all sites but $k$. The post-measured state is $\tilde{\rho}_{m,k}^{(g)} \!=\!K_{m,k}^{(g,g)} \rho_k K_{m,k}^{(g,g)\dagger}/p_{m,k}^{(g)}$, with $p_{m,k}^{(g)}\!=\!\text{Tr}(K_{m,k}^{(g,g)} \rho_k K_{m,k}^{(g,g)\dagger})$, is probability of clicking $m$ at site $k$.
For even $k$, expanding Kraus operators $K_{m,k}^{(1,1)}$ in Eq.~(\ref{eq:def_K}) for $m\!=\!0,1$ at infinitesimal $\epsilon>0$ gives
\begin{align}
    \tilde{\rho}_{0,k}^{(1)} &= \rho_k-\frac{\epsilon^2}{2}\left\{\hat{n}_k^{(1)}-\langle\hat{n}^{(1)}_k\rangle,\rho_k\right\},& \tilde{\rho}_{1,k}^{(1)} &= \ket{1}\!\bra{1}_k \\
    \nonumber
    p_{0,k}^{(1)} &= 1-p_{1,k}^{(1)},& p_{1,k}^{(1)} &= \langle\hat{n}^{(1)}_{k}\rangle\epsilon^2,
\end{align}
where $\{A,B\}\!=\!AB+BA$ is the anti-commutator. For infinitesimal $\epsilon$ at time step $dt$ with $\epsilon^2=h_2^{\prime} dt$ for weak continuous measurement of strength $h_2^{\prime}$, the probability of clicking $m\!=\!1$, $p_{1,k}^{(1)}\approx\langle\hat{n}^{(1)}_k\rangle h_2^{\prime} dt\ll1$. Therefore, with continuous clicks of $m\!=\!0$, i.e., post-selecting the trajectory of maximum probability, the effective action is obtained with
\begin{align}
    d\rho_k=\tilde{\rho}_{0,k}^{(1)}-\rho_k = -\frac{h_2^{\prime}}{2}dt\left\{\hat{\sigma}^z_k-\langle\hat{\sigma}^z_k\rangle, \rho_k\right\}.
\end{align}
Similarly, for odd sites, i.e., $g\!=\!0$, $p_{0,k}^{(0)}\approx\langle\hat{n}^{(0)}_k\rangle h_2^{\prime} dt\ll1$, with $m\!=\!1$ giving the post-selection and 
\begin{align}
    d\rho_k=\tilde{\rho}_{1,k}^{(0)}-\rho_k = \frac{h_2^{\prime}}{2}dt\left\{\hat{\sigma}^z_k-\langle\hat{\sigma}^z_k\rangle, \rho_k\right\},
\end{align}
giving the alternating effect of $h_2^{\prime}$ on odd and even sites.

For the $N$-qubit state $\rho$, evolving under Hermitian Hamiltonian $\hat{\mathcal{H}}_1$, interacting with continuously monitored bath which under post-selection leads to a non-Hermitian Hamiltonian,
\begin{align}
    &\hat{\mathcal{H}}_{nH} = \hat{\mathcal{H}}_1 + i\hat{\mathcal{H}}_2;\quad \hat{\mathcal{H}}_2 = h_2^{\prime}\sum_{k=1}^N (-1)^{k+1} \hat{\sigma}_k^z
\end{align}
The effective dynamics follows
\begin{align}
    \nonumber
    &\frac{d\rho}{dt} \!=\! -i\!\left(\hat{\mathcal{H}}_{nH}\rho \!-\!\rho\hat{\mathcal{H}}_{nH}^{\dagger}\!\right) \!\Rightarrow \rho(t) \!=\! \frac{e^{-i\hat{\mathcal{H}}_{nH}t}\rho(0)e^{i\hat{\mathcal{H}}_{nH}^{\dagger}t}}{\text{Tr}\!\left(\!e^{-i\hat{\mathcal{H}}_{nH}t}\rho(0)e^{i\hat{\mathcal{H}}_{nH}^{\dagger}t}\!\right)}
\end{align}
with non-Hermitian Hamiltonian $\hat{\mathcal{H}}_{nH}$.

\section{Fidelity for quantum state transfer with $U(1)$-symmetric non-Hermitian Hamiltonian}
\label{app:fid_deriv}

A $U(1)$-symmetric (non-Hermitian) Hamiltonian,  \(\hat{\mathcal{H}}\), can be simplified in the sectors of conserved charges, i.e., conserved total magnetization $\hat{S}\!=\!\sum_{k=1}^{N}\hat{\sigma}^z_k$. Therefore, \(|0\rangle^{\otimes N}\) is an eigenstate of the system with eigenvalue $e_0$, whereas 
$\ket{\mathbf{k}} $ are $N$ orthogonal states with $k\!=\!1$ $,...,$ $N$. $\{\ket{\mathbf{k}}\}_k \!=\! \ket{1}_k\!\!\bigotimes\limits_{j=1,j\neq k}^{N}\!\ket{0}_{j}$ is an $N$-dimensional eigenspace of the Hamiltonian. Setting $e_0\!=\!0$ (by $\hat{\mathcal{H}}\to\hat{\mathcal{H}}-e_0\hat{\mathbb{I}}$), the time-evolved state of the overall system with effective evolution of \(|1\rangle_1\otimes|0\rangle_{2,3,...,N} \equiv |\textbf{A}\rangle\) term only, is given by
\begin{equation}
|\Psi'(t)\rangle = \cos\!\frac{\theta}{2}|0\rangle^{\otimes N}
 + e^{i\phi}\sin\!\frac{\theta}{2}
 \sum_{k=1}^N
 \langle \mathbf{k}| e^{-i\hat{\mathcal{H}}t} |\mathbf{A}\rangle
 |\mathbf{k}\rangle,
\end{equation}
for the sender, Alice, $\ket{\mathbf{A}}\!\equiv\!\ket{\mathbf{1}}$ at site $1$ and the receiver, Bob, $\ket{\mathbf{B}}\!\equiv\!\ket{\mathbf{N}}$ at site $N$.

When the Hamiltonian is non-Hermitian, the overall state has to be normalised at every step of time evolution~\cite{Lee2014, Tu2023}, so the normalization constant becomes
\begin{align}
\nonumber \mathcal{N} & \equiv \langle \Psi'(t)|\Psi'(t)\rangle  \\ \nonumber  &= \cos^2\frac{\theta}{2} + \sin^2\frac{\theta}{2} \sum_{k=1}^N \langle s|e^{i\hat{\mathcal{H}}^\dagger t} |\textbf{k}\rangle \langle \textbf{k} |e^{-i\hat{\mathcal{H}} t}|s \rangle \\
& = \cos^2\frac{\theta}{2} + \sin^2\frac{\theta}{2} \sum_{k=1}^N |b_{k}|^2.
\end{align}

Hence, the state after evolution is
\begin{eqnarray}
|\Psi(t)\rangle = |\Psi'(t)\rangle/\sqrt{\mathcal{N}}.
\end{eqnarray}
Now, at a given time $t$, tracing out the \((N-1)\) spins gives the state at the Bob's  site (\(N\)) as
\begin{align}
\nonumber \rho_{N}(t) &= (1 - \sin^2\!\frac{\theta}{2}\frac{|b_B|^2}{\mathcal{N}})|0\rangle \langle0| + \nonumber e^{i\phi} \sin\!\frac{\theta}{2} \cos\!\frac{\theta}{2}\frac{b_B}{\mathcal{N}}|1\rangle \langle0| \\ &+ \nonumber e^{-i\phi} \sin\frac{\theta}{2} \cos\frac{\theta}{2}\frac{b_B^*}{\mathcal{N}}|0\rangle \langle1| + \sin^2\frac{\theta}{2}\frac{|b_B|^2}{\mathcal{N}} |1\rangle \langle1|
\end{align}
where \(b_B = \langle \textbf{B}|e^{-i \hat{\mathcal{H}t}}|\textbf{A}\rangle\) represents the amplitude of transition of the Alice's state to the Bob's site through the time evolution. The fidelity, which quantifies the overlap between the initial arbitrary state at Alice's end, and the state received by Bob at the end of the protocol after the time evolution, is given as
\begin{eqnarray}
\mathcal{F}_N(t) &= &\int_{\text{Haar}} \langle \psi|\rho_N(t)|\psi\rangle d|\psi\rangle \nonumber \\
&=& \frac{1}{4 \pi} \int_{\theta =0}^{\pi} \int_{\phi=0}^{2\pi}  \langle \psi|\rho_N(t)|\psi\rangle \sin \theta d\theta d\phi.
\label{eq:fid}
\end{eqnarray} 
Here, Haar average is taken over all possible states corresponding to different values of the parameters \((\theta, \phi)\).
By computing the terms in our case, using the above equation, the final expression becomes
\begin{align}
\nonumber \mathcal{F}_N(t) &= \frac{1}{2} + \frac{(|b_B|\cos\gamma - 2|b_B|^2 + \Gamma|b_B|\cos\gamma)}{(\Gamma-1)^2} \\ & + \frac{((1+\Gamma)|b_B|^2 - 2\Gamma|b_B|\cos\gamma) \log \Gamma}{(\Gamma-1)^3},
\end{align} 
where \(|b_B|\cos\gamma\) is the real part of \(b_B\) and \(\Gamma=\sum_{k=1}^N |b_{k}|^2\).



\section {Expression of fidelity for $XX$ model with \(N=2\)}
\label{sec:appen_xx}

The $XX$ Hamiltonian with alternating field at neighboring sites for \(N=2\) becomes
\begin{eqnarray}
 \nonumber \hat{\mathcal{H}}_{nH} =  \hat{\sigma}^x_1\hat{\sigma}^x_{2} + \hat{\sigma}^y_1\hat{\sigma}^y_{2} +  (h_1 + ih_2)\hat{\sigma}_1^z + (h_1 - ih_2)\hat{\sigma}_2^z.\qquad
\end{eqnarray} 
The corresponding set of eigenvalues is (\(-2h_1, -2\sqrt{1 - h_2^2},2\sqrt{1 - h_2^2}, 2h_1 \)), where taking \(\sqrt{1 - h_2^2} \equiv \mathfrak{J}\), and scaling the eigenvalues, the final set becomes (\(0, -2\mathfrak{J} +2h_1, 2\mathfrak{J} +2h_1, 4h_1 \)), since the eigenstate of the Hamiltonian must be \(|00\rangle\). Thus broken phase emerge when \(h_2 > 1\) and is unbroken when \(h_2 < 1\). Now, we need to calculate \(|b_B|\) and \(\Gamma\) for both the cases in order to get the corresponding expressions of fidelities, where
\begin{eqnarray}
b_B = \langle \textbf{B} | e^{-i\hat{\mathcal{H}}t}|\textbf{A}\rangle, && \\ 
 |\textbf{B}\rangle = |01\rangle,  |\textbf{A}\rangle = |10\rangle.
\end{eqnarray} 
Since \(e^{-i\hat{\mathcal{H}}t} = Pe^{-iEt}P^{-1}\), \(E\) is the diagonal matrix of the eigenvalues given as diag(\(0, -2\mathfrak{J} +2h_1, 2\mathfrak{J} +2h_1, 4h_1 \)) and \(P\) is the matrix corresponding to the eigenvectors, which reads as
    \begin{equation}
    P=\left[
\begin{array}{cccc}
  0 & 0 & 0 & 1 \\
0 & (ih_2 - \mathfrak{J}) & (ih_2 + \mathfrak{J}) & 0 \\
 0 &  1 & 1 & 0 \\
  1 & 0 & 0 & 0 \\
\end{array}
\right ].
\end{equation}
Hence, 
\begin{eqnarray}
b_B = -\frac{\sinh(2it\mathfrak{J})}{\mathfrak{J}}e^{-2ith_1}.
\end{eqnarray}
Now, \(\Gamma = |\langle 01|e^{-i\hat{\mathcal{H}}t}|10\rangle|^2 + |\langle 10|e^{-i\hat{\mathcal{H}}t}|10\rangle|^2\), its first term is \(|b_B|^2\), while its second term can be obtained from
\begin{equation}
\langle 10|e^{-i\hat{\mathcal{H}}t}|10\rangle = \!\left(\cosh(2it\mathfrak{J}) +ih_2\frac{\sinh(2it\mathfrak{J})}{\mathfrak{J}}\right)e^{-2ith_1}.
\end{equation}
First, considering the unbroken regime, i.e., when \(h_2 < 1\):
\begin{eqnarray}
\nonumber \Gamma = \cos^2(2t\mathfrak{J})) + h_2\frac{\sin(4t\mathfrak{J})}{\mathfrak{J}} + \frac{\mathfrak{J}^2 + 2h_2^2}{\mathfrak{J}^2}\sin^2(2t\mathfrak{J})).
\end{eqnarray}
From the above expression, we understand that the cosine and sine terms lead to oscillations and all the terms in the Eq. (\ref{eq:fid_eqn}) give significant contribution for some values of the system parameters.

For broken regime, i.e., when \(h_2 > 1\) and taking \(\sqrt{h_2^2 - 1} \equiv \tilde{\mathfrak{J}}\) where \(\tilde{\mathfrak{J}}\) is real, we obtain
\begin{eqnarray}
\nonumber \Gamma = \cosh^2(2t\tilde{\mathfrak{J}}) + h_2\frac{\sinh(4t\tilde{\mathfrak{J}})}{\tilde{\mathfrak{J}}} +  \frac{\tilde{\mathfrak{J}}^2 + 2h_2^2}{\tilde{\mathfrak{J}}}\sinh^2(2t\tilde{\mathfrak{J}}).
\end{eqnarray}
Here in the broken regime, we understand that the hyperbolic sine and cosine terms lead to exponential rise in \(\Gamma\), which ultimately leads to exponential decrease in the fidelity and hence the fidelity does not get any addition from the second and third terms of Eq. (\ref{eq:fid_eqn}) and remains almost stagnated at \(0.5\) value.

\bibliography{ref.bib}


\end{document}